\documentclass[amsmath,amssymb,twocolumn,prb]{revtex4}
\usepackage{amsmath}
\usepackage{amssymb}
\usepackage{graphicx}
\usepackage{bm}
\usepackage{dcolumn}

\newcommand{\bpm}{\begin{pmatrix}}
\newcommand{\epm}{\end{pmatrix}}
\newcommand{\bfk}{{\boldsymbol{k}}}
\newcommand{\bfB}{{\boldsymbol{B}}}

\newcommand{\eps}{\varepsilon}

\begin{document}

\title{Antiferromagnetism and hot spots in CeIn$_3$}

\author{L. P. Gor'kov}
\author{P. D. Grigoriev}
 \email{grigorev@magnet.fsu.edu}
\altaffiliation[Also at ]{L. D. Landau Institute for Theoretical
Physics,
 Chernogolovka, Russia}
\affiliation{ National High Magnetic Field laboratory, Florida State
University, Tallahassee, Florida}
\date{\today }

\begin{abstract}

Enormous mass enhancement at ''hot spots'' on the Fermi surface (FS)
of CeIn$_3$ has been reported at strong magnetic field near its
antiferromagnetic (AFM) quantum critical point [T. Ebihara et al.,
Phys. Rev. Lett. 93, 246401 (2004)] and ascribed to anomalous spin
fluctuations at these spots. The ''hot spots'' lie at the positions
on FS where in non-magnetic LaIn$_3$ the narrow necks are protruded.
In paramagnetic phase CeIn$_3$ has similar spectrum. We show that in
the presence of AFM ordering its FS undergoes a topological change
at the onset of AFM order that truncates the necks at the ''hot
spots'' for one of the branches. Applied field leads to the
logarithmic divergence of the dHvA effective mass when the electron
trajectory passes near or through the neck positions. This effect
explains the observed dHvA mass enhancement at the ''hot spots'' and
leads to interesting predictions concerning the spin-dependence of
the effective electron mass. The (T,B)-phase diagram of CeIn$_3$,
constructed in terms of the Landau functional, is in agreement with
experiment.

\end{abstract}

\maketitle

{\it Introduction.} Interest to the phenomena at quantum critical
point (QCP) pervades the current
literature\cite{Hertz,Millis,Mathur,Paschen} on intermetallic
compounds. Recently the effect of magnetic fields has been studied
in the antiferromagnetic (AFM) CeIn$_3$.\cite{HarrCeIn3} The {\it
magnetic} QCP was found at $B_c= 61T$. A strong mass enhancement
observed via  the de Haas - van Alphen (dHvA) effect for the
electron trajectories that cross or pass close to some ''hot'' spots
at the Fermi surface (FS), has been reported and interpreted in
terms of strong spin fluctuations at ''hot'' spots implying strong
many-body interactions.\cite{HarrCeIn3} It was noted that positions
of these ''hot'' spots coincide with the positions of the necks
protruding from the similar FS for non-magnetic LaIn$_3$. The necks
would fall close to the boundary of the AFM Brillouin Zone (BZ) for
CeIn$_3$ and must be somehow changed or even truncated due to
electron reflections at the new BZ. The AFM propagation vector
$\boldsymbol{Q}=(\pi/a)(1,1,1)$ connects opposite spots on the Fermi
surface.\cite{Knafo} Topological changes of the FS geometry near
necks, known as the Lifshitz ''2.5''-transitions, lead to weak
singularities in thermodynamic and transport properties.\cite{LAK}
We show that at proper field directions this also affects the dHvA
characteristics at ''hot'' spots.

We consider dHvA effect for the three field orientations in the {\it
cubic} CeIn$_3$ or LaIn$_3$. The mass enhancement was reported for
two field orientations: $\boldsymbol{B}\parallel (110)$ and
$\boldsymbol{B}\parallel (111)$.\cite{HarrCeIn3} In the first case
the extremal electron trajectory would run across the four necks on
the FS. More detailed measurements in Ref. [\onlinecite{HarrCeIn3}]
were performed for $\boldsymbol{B}\parallel (111)$. In this case the
extremal orbit does not cross "hot" spots, but may run close if the
necks` diameter is large enough. The dHvA measurement at
$\boldsymbol{B}\parallel (100)$ when the electron trajectory passes
far away from any ''hot'' spot provides the non-enhanced effective
mass value $m_1=2m_0$.\cite{HarrCeIn3}

\begin{figure}[tb]
\includegraphics{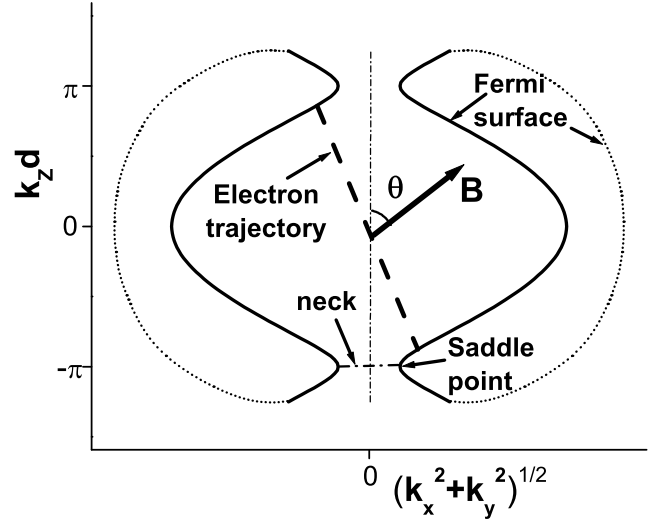}
\caption{\label{FSFig} The Fermi surface (solid line) and the
extremal electron orbit (dashed line). The saddle points due to
necks are shown. The dotted lines schematically show the outer Fermi
surface in CeIn$_3$ (see
[\onlinecite{FSCeIn3}],[\onlinecite{FSCeIn3N}]).}
\end{figure}

To start with, we explain in frameworks of a simple model, the
larger masses for $\boldsymbol{B}\parallel (111)$ by the electron
trajectory proximity to the saddle points, where the dHvA effective
mass has a logarithmic singularity.\cite{LAK} If the FS has necks,
at certain orientations of the magnetic field the extremal
cross-section electron trajectory passes through a saddle point (see
Fig. \ref{FSFig}).

{\it Model calculations.} We take a Fermi surface with axial
symmetry along $z$-axis and narrow necks as shown in Fig.
\ref{FSFig}. For brevity we omit initially the effects of the AFM
ordering and other spin effects on the ''bare'' electron dispersion
chosen as (we set $\hbar =1$)
  \begin{equation}
   \label{dispersion1}
   \eps (\bfk ) = (k_x^2+k_y^2)/2m_1 +2t_{z}[1-\cos(k_z d)],
 \end{equation}
where $k_{x,y,z}$ are the momentum components along $x,y,z$ axes and
$t_{z}$ is the transfer integral along $z$-direction. $k_zd=\pm\pi$
defines the neck positions. The Fermi surface ($\eps (\bfk
)=\eps_F$) is given by
  \begin{equation}
   \label{FS}
   (k_x^2+k_y^2)/2m_1 -2t_{z}[1+\cos(k_z d)]=\Delta ,
 \end{equation}
where $\Delta\equiv\eps_F - 4t_z$. The neck radius is $
k_{neck}=\sqrt{2m_1\Delta}$. In CeIn$_3$ and LaIn$_3$ the necks are
narrow, $k_{neck}d/\pi\ll 1$.

The magnetic field $\bfB = B_0(sin \theta , 0, \cos\theta )$ in Fig.
\ref{FSFig} directed at angle $\theta =70.53^{\circ}$ with respect
to the $z$-axis would simulate $B\parallel (111)$ in the cubic
CeIn$_3$ where the extremal orbits, as we shall see, run rather
close to the {\it other six} ''hot'' spots but do not cross
them.\cite{HarrCeIn3} Rotate the $x$-$z$ coordinate axes by the
angle $\theta$ to make the direction of magnetic field along the
$k_z'$-axis:
  \begin{equation}
   \label{newaxes}
    \left\{
   \begin{array}{c}
k_x=k_x'\cos\theta +k_z'\sin\theta \\
k_z=-k_x'\sin\theta +k_z'\cos \theta .
   \end{array} \right.
 \end{equation}
The dispersion relation (\ref{dispersion1}) in the new variables is:
  \begin{equation}
   \label{dispersion2}
\begin{split}
   \eps (\bfk ') &= [(k_x'\cos\theta +k_z'\sin\theta
)^2+k_y'^2]/2m_1\\
   &+2t_{z}\{1-\cos [d(-k_x'\sin\theta +k_z'\cos \theta )]\}.
\end{split}
 \end{equation}
Electrons move along the quasi-classical trajectories of constant
energy and constant $k_z'$. In the dHvA effect only the electron
trajectories which encircle the extremal cross-sections
$S_{extr}(k_z'=0)$ of the FS are important.

The effective electron mass is determined by\cite{LAK}
  \begin{equation}
   \label{m*}
2\pi m^*=\frac{\partial S_{extr}}{\partial \eps }=\oint
\frac{dk_{\perp}}{v_{\perp}}.
 \end{equation}
Along the trajectory of constant $\eps (\bfk '),k_z'$, $d\eps
(k_x',k_y')=v_x' dk_x'+v_y'dk_y'=0$, and the integral (\ref{m*})
rewrites as
  \begin{equation}
   \label{m*1}
2\pi m^*(\theta )=\oint \frac{dk_x'}{v_{y}'}= 4\int_0^{k_{x0}}
\frac{dk_x'}{v'_y(k_x')},
 \end{equation}
where
  \begin{equation}
   \label{vy}
v'_y(k_x')=\frac{\sqrt{2m_1[\Delta+2t_{z}[1+\cos(k_x'd\sin\theta)]]-k_x
^{'2}\cos^2\theta }}{m_1}
 \end{equation}
and the integration limit $k_{x0}(\theta )$ near the necks is the
solution of equation $v'_y(k_{x0}',\theta )=0$. The extremal
electron trajectory passes through the saddle point if in addition
to the Eqs. (\ref{vy}), the condition
$v_{x}(\theta)\vert_{k'_y=k'_z=0,\eps =\eps_F}=0$ is satisfied. From
(\ref{vy}) one finds the critical tilt angle $\theta_c$ (for narrow
necks $\tan \theta_c \approx\pi/k_{neck}d$) corresponding to a jump
of the dHvA frequency. For lower $\theta$ the electron trajectory
does not pass through the saddle point. At higher tilt angles
$\theta$ the electron trajectory over the necks comes to the next
BZ.

Consider the case of a narrow neck. Introducing
$\overline{\theta}=\pi /2-\theta$ one get for the saddle point,
$\tan\overline{\theta}_c=k_{neck}d/2\pi \ll 1$. Taking $k'_z=0$ in
Eq. (\ref{newaxes}), return in (\ref{m*1}) to the integration over
$k_z$: $dk'_x=-dk_z/\cos\overline{\theta}$. Expanding $v'_y(k_x')$
(\ref{vy}) near $k'_{x0}$ at $\overline{\theta}$ close to
$\overline{\theta}_c$, one obtains ($\delta\equiv\pi /d-k_z\ll 1$
and $k_x=-k_z\tan\overline{\theta}$)
  \begin{equation}
   \label{m*narrowneck}
m^*(\overline{\theta})= \frac{2}{\pi d}
\left(\frac{m_1}{t_z}\right)^{1/2}\int_{\delta_0}^{\sim
\frac{\pi}{d}} \frac{d\delta} {\sqrt{
\delta^2+\left(\frac{\pi}{d}\right)^2
\left(\frac{\overline{\theta}_c^2-\overline{\theta}^2}{m_1t_zd^2}\right)}},
 \end{equation}
 where $\delta_0\equiv \left(\pi /d \right)\sqrt{
(\overline{\theta}^2-\overline{\theta}_c^2) /m_1t_zd^2}$, i.e.
$m^*(\overline{\theta})$ is logarithmically divergent as
  \begin{equation}
   \label{LogDiv}
m^*(\overline{\theta})= (2/\pi)\left(m_1/d^2 t_z\right)^{1/2}
\ln\left(1/d\delta_0\right) .
 \end{equation}

For $\boldsymbol{B}\parallel z$-axis in Fig.1 the dHvA oscillations
in the model of Eq. (\ref{dispersion1}) only measure the central
(''belly'') cross section and the thickness of the ''neck''. Should
we return to the cubic case and $\boldsymbol{B}\parallel (111)$, the
extremal trajectories cross none of the six ''hot'' spots in
[[\onlinecite{HarrCeIn3}],Fig.3] but numerically run rather close to
them (the deviation of the trajectory from the center of the ''hot
spot'' is given by $\overline{\theta}=19.47^{\circ}$). Thus, it
becomes a question how broad are the necks to lead to a significant
mass enhancement.

From the dHvA data on LaIn$_3$ (Fig. 4 of Ref. [\onlinecite{LaIn3}])
one knows the neck and ''belly'' cross-section areas: $k_{neck}
a=0.27$ and (for the spherical FS denoted as (d) \cite{Handbook})
$k_{F} a \approx 2.24$. Taking the value $d=2a/\sqrt{3}$ in Eq.
(\ref{dispersion1}), as for the cubic lattice, this gives $ k_{neck}
d\approx 0.32$ and $k_{F} d\approx 2.6$. The saddle points in
LaIn$_3$ appear at the angle $\overline{\theta}_c\approx
5.8^{\circ}$. Hence, the necks in nonmagnetic LaIn$_3$ are too
narrow to affect the dHvA effective mass for
$\boldsymbol{B}\parallel (111)$.

In CeIn$_3$ the ''belly'' radius $k_F$ is very close to the one in
LaIn$_3$, while the neck cross-section area depends on the AFM order
parameter and the value of magnetic field (in addition, the bands
become spin split, see below). At field $B=15T$, the dHvA frequency
from the neck (the (j)-orbit) is about $3$ times larger than in
LaIn$_3$,\cite{FSCeIn3} which would give $k_{neck} d\approx 0.55$
and $\overline{\theta}_c\approx 10^{\circ}$. It is still rather far
from the tilt angle $\overline{\theta}_0=19.47^{\circ}$. However,
according to Refs. [\onlinecite{HarrCeIn3,FSCeIn3N}] the neck radius
is considerably higher. Then the neck radius may reach and overpass
the critical value $k_{neck}^{crit}d\approx 1.1$ when
$\overline{\theta}_c=\overline{\theta}_0$ for the extremal orbit to
pass through the saddle point at the field $\boldsymbol{B}\parallel
(111)$. There are no data on the field dependence of the neck radius
in CeIn$_3$ so far.

If $\boldsymbol{B}$ is perpendicular to the plane in Fig.1 for our
model (\ref{dispersion1}), the extremal trajectory would run along
the FS shown in Fig.1. The mass enhancement would be determined by
the neck`s width $m^*\approx 2m_1\sqrt{2\hbar^2/t_zm_1d^2} \ln
(2\pi\sqrt{t_z/\Delta})$. In the cubic CeIn$_3$ this would
correspond to $\boldsymbol{B}\parallel (110)$: four necks`
singularities would provide strong mass enhancement, as stated in
Ref. [\onlinecite{HarrCeIn3}]. Strictly speaking, the dHvA frequency
for the (d)-FS should be observable only at strong band splitting,
as we discuss below.

{\it CeIn$_3$ and other REIn$_3$.}  The FS`es in CeIn$_3$ are now
known in some main details.\cite{HarrCeIn3,FSCeIn3,FSCeIn3N} The two
most remarkable features are common with nonmagnetic
LaIn$_3$\cite{LaIn3}: 1)a practically spherical FS sheet (denoted as
(d)\cite{Handbook,FSCeIn3,FSCeIn3N,Umehara}) with the diameter in
the k-space close to the AFM vector $\boldsymbol{Q}=(\pi
/a)(1,1,1)$; 2) ''necks'' protruding from FS sheet (d) towards an
outer FS.\cite{FSCeIn3,FSCeIn3N} The dHvA orbits and FS`s for necks
were labeled as (j); their sizes vary among the REIn$_3$
group.\cite{Handbook,Umehara} Analysis of the dHvA oscillations, as
well as the electron-positron annihilation experiments in the
paramagnetic phase\cite{FSCeIn3,FSCeIn3N} confirm the localized
character of the Ce f-electrons.\cite{Knafo,Walker,CePd2Si2} The
moment J=5/2 in the cubic environment is split into the quartet,
$\Gamma_8$ and the Kramers` doublet $\Gamma_7$; the latter is
responsible for the AFM ordering in CeIn$_3$ \cite{Knafo}. The
propagating vector $\boldsymbol{Q}$ corresponds to the staggered
magnetization, $\boldsymbol{S}_{\perp}(\boldsymbol{Q})$, aligned
antiferromagnetically perpendicular to the adjacent $(111)$
ferromagnetic planes.\cite{Walker} Magnetic anisotropy seems to be
weak\cite{CePd2Si2} and is neglected below.

CeIn$_3$ is a moderate HF material with the Sommerfeld`s $\gamma
=130mJ/K^2$mole. At the ambient pressure the Neel temperature is
$T_N=10K$. The staggered magnetization is close to the value
0.71$\mu_B$ expected for localized $\Gamma_7$ doublet (see in Ref.
[\onlinecite{HarrCeIn3}]). The AFM state can be suppressed by
applied pressure $P_c\approx 26$ kbar.\cite{Endo} In the vicinity of
this pressure the coexistence of AFM order and superconductivity has
been reported.\cite{Mathur} The {\it magnetic} QCP in CeIn$_3$ was
found at the field $B\approx 61T$.\cite{HarrCeIn3} As it was said
above, the authors claimed {\it strong many-body effects} at ''hot''
spots on the FS sheet (d).\cite{HarrCeIn3}

{\it AFM order in CeIn$_3$.} For common antiferromagnets strong
enough applied field destroys the Neel state by aligning staggered
moments parallel to the field, $\boldsymbol{B}$. For CeIn$_3$
$T_N=10K$ looks already rather low, so that one may attempt apply
the Landau mean-field approach:\cite{LL1}
  \begin{equation}
   \label{LandauExpansion}
F(T,B)= a(T-T_N)\boldsymbol{S}_{\perp}^2 + b
\boldsymbol{S}_{\perp}^4-\frac{\chi \boldsymbol{B}^2}{2}+\eta
(\boldsymbol{B}\boldsymbol{S}_{\perp})^2+\eta'\boldsymbol{B}^2\boldsymbol{S}_{\perp}^2
  \end{equation}
where $\boldsymbol{S}_{\perp}$ is the local spin component along the
staggered magnetization vector (only terms independent on the
crystal anisotropy left in Eq. (\ref{LandauExpansion})). From Eq.
(\ref{LandauExpansion}) the quadratic dependence
$T_N(B)=T_N[1-(B/B_c )^2]$ immediately follows, reproducing the
results on Fig.1a of Ref. [\onlinecite{HarrCeIn3}] with high
accuracy. This agrees with the assumption that magnetic anisotropy
is indeed low. The dHvA mass for the (d) sheet away from the
singular field orientations is also rather low: $m^*\approx
2m_e$.\cite{HarrCeIn3} Therefore, we assume that AFM order and the
phenomena studied in Ref. [\onlinecite{HarrCeIn3}] are only weakly
linked to other Kondo-like features of CeIn$_3$: i.e., the (d) and
(j) FS pieces are weakly coupled to the f-electrons. Next question
is whether one can explain low $T_N =10K$ for CeIn$_3$ via the RKKY
mechanism with the help of (d)-sheet only. The diameter of the
(d)-sheet equals to the $\boldsymbol{Q}$ vector value and, hence, is
capable to provide the commensurate RKKY interaction. It is also
known that parameters for these FS`es are not dramatically different
for the rest of REIn$_3$ family.\cite{Handbook} From Eq.
(\ref{LandauExpansion}) it follows that $
|\boldsymbol{S}_{\perp}|(\boldsymbol{Q})\approx (1-T/T_N)^{1/2}$
,(at $B=0$), and $|\boldsymbol{S}_{\perp}|=(1-(B/B_c)^2)^{1/2}$ at
$T=0$.

{\it Energy spectrum near hot spots at non-zero $B$ and
$|\boldsymbol{S}_{\perp}|$.} Introduce the exchange term
$J\hat{\boldsymbol{\sigma}}\boldsymbol{S}_i$ between itinerant and
localized spins. This exchange leads to the RKKY interaction between
localized spins. It can be estimated as $J\sim\sqrt{T_N/\nu_F}$,
where $\nu_F$ is the density of states at the Fermi level. Assuming
that only the (d)-parts of electron spectrum contribute to the RKKY
interaction we obtain $\nu_F\approx 3/2\eps_F\approx
3m_1/k_{belly}^2=1/3000K$. This gives $J\approx 170K$. Therefore, at
all $\boldsymbol{B}$, $\mu_BB\ll J$.

\begin{figure}[tb]
\includegraphics{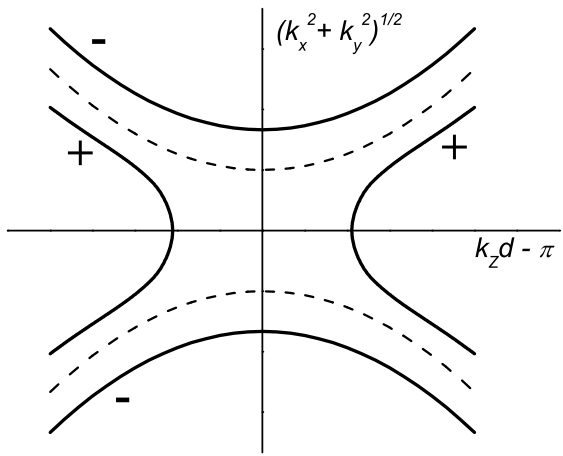}
\caption{\label{FS2Fig} Schematic view of the Fermi surface at the
neck with (solid line) and without (dashed line) AFM order. The
($\pm$) signs on the figure correspond to the ($\pm$) signs in Eq.
(\ref{Eperp})}
\end{figure}

The effective magnetic field acting on electrons is
  \begin{equation}
   \label{Beff}
\mu_{B}\boldsymbol{B}_{eff}=\mu_B\boldsymbol{B}+J\boldsymbol{S}_{\parallel},
\end{equation}
where $\boldsymbol{S}_{\parallel}$ is the local spins' component
parallel to $\boldsymbol{B}$. Let $\hat{\boldsymbol{n}}_{S}$ and
$\hat{\boldsymbol{n}}_{B}$ be the two perpendicular (in absence of
anisotropy) unit vectors of Pauli matrices for the directions of
$\boldsymbol{S}_{\perp}$ and $\boldsymbol{B}$ correspondingly. Then
with $\boldsymbol{Q}$ being the AFM propagation vector, the new
energy spectrum in AFM phase is determined from two equations for
the electronic states $(\boldsymbol{k},\boldsymbol{k+Q})$:
  \begin{equation}
   \label{EqForEnSp}
\left\{   \begin{array}{c}
(\hat{E}-\hat{\eps}_{\boldsymbol{k}}+\hat{\boldsymbol{n}}_{B}\mu_{B}\boldsymbol{B}_{eff})\Psi_{\boldsymbol{k}}
=-J \hat{\boldsymbol{n}}_{S} \boldsymbol{S}_{\perp} \Psi_{\boldsymbol{k}+\boldsymbol{Q}}  \\
(\hat{E}-\hat{\eps}_{\boldsymbol{k}+\boldsymbol{Q}}+\hat{\boldsymbol{n}}_{B}\mu_{B}\boldsymbol{B}_{eff})
\Psi_{\boldsymbol{k}+\boldsymbol{Q}} =-J \hat{\boldsymbol{n}}_{S}
\boldsymbol{S}_{\perp} \Psi_{\boldsymbol{k}} .
  \end{array} \right.
 \end{equation}
Multiplying both equations by $\hat{\boldsymbol{n}}_{S}$ and
excluding $\Psi_{\boldsymbol{k}+\boldsymbol{Q}}$ one obtains the
equation in the spin space:
  \begin{equation}
   \label{EqForEnSpt}
 \begin{array}{c}
(\hat{E}-\hat{\eps}_{\boldsymbol{k}+\boldsymbol{Q}}-\hat{\boldsymbol{n}}_{B}
\mu_{B}\boldsymbol{B}_{eff})
\\
\times
(\hat{E}-\hat{\eps}_{\boldsymbol{k}}+\hat{\boldsymbol{n}}_{B}\mu_{B}\boldsymbol{B}_{eff})
\Psi_{\boldsymbol{k}} =J ^2 |\boldsymbol{S}_{\perp}|^2
\Psi_{\boldsymbol{k}}
 \end{array}
 \end{equation}
The four energy branches from (\ref{EqForEnSpt}) are
\begin{equation}
   \label{Eperp}
E_{k,\sigma}^{\pm} =\frac{\eps_{k}+\eps_{k+Q}}{2}\pm \sqrt{\left(
\frac{\eps_{k}-\eps_{k+Q}}{2}-\mu_B
B_{eff}\sigma\right)^2+J^2S_{\perp}^2}.
\end{equation}
If the vector $\boldsymbol{Q}$ exactly connects two opposite necks
in Fig. \ref{FSFig}, one obtains near the necks
\begin{equation}
   \label{Esimple}
E_{k,\sigma}^{\pm} =\eps_{k}\pm \sqrt{\left( \mu_B B_{eff}
\right)^2+J^2|S_{\perp}|^2}.
\end{equation}
Above $B_c$ the two branches go over into the Zeeman splitting with
the effective field from Eq. (\ref{Beff}) and $S_{\parallel}=1$. For
convenience, we normalize
$|\boldsymbol{S}_{\parallel}|^2+|\boldsymbol{S}_{\perp}|^2=1$. Then
substituting Eq. (\ref{Beff}) to (\ref{Esimple}) we obtain at
$\mu_BB\ll J$
\begin{equation}
   \label{EB}
E_{k,\sigma}^{\pm} \approx \eps_{k}\pm (J+\mu_B\boldsymbol{B}\,
\boldsymbol{S}_{\parallel}).
\end{equation}
The AFM order adds new features to the electron dispersion and the
Fermi surface of the model (\ref{dispersion1}). With signs ($\pm $)
there are now two branches shown in Fig. \ref{FS2Fig}. For the sign
(+) the dHvA experiment at $B\ll B_c$ would see the decrease of the
necks' width as compared to the bare spectrum (i.e., without AFM
order). If $\Delta =k_{neck} ^2/2m_1 <J$, these necks in the
(d)-part of the FS are completely destroyed, as shown in Fig. 2.
Otherwise, these necks would only be narrowed. For the sign (-) the
necks' width increases due to AFM ordering.

The authors of Ref. [\onlinecite{HarrCeIn3}] have used the width of
the necks $\approx 1/12$ of the orbit circumference. This is close
to the data of Ref. [\onlinecite{FSCeIn3N}] but disagree with the
estimations of Ref. [\onlinecite{FSCeIn3}]. Using data of Ref.
[\onlinecite{FSCeIn3N}] we would get $\Delta$ comparable to $J$. If
so, one may indeed relate the mass enhancement mechanism for
$\boldsymbol{B}\parallel (111)$ described above to the broadening of
the gap between two outer (-) trajectories in Fig. \ref{FS2Fig}.
This broadening depends on the magnetic field as described by Eq.
(\ref{EB}). At $J=170K$ and $B=60T$ this term is $\approx 0.4J$. The
estimates we have done for $\Delta$,$J$ show that it is realistic to
account for the observed mass enhancement at
$\boldsymbol{B}\parallel (111)$. On the other hand, we must repeat
that there are no experimental data for a quantitative fit because
so far no attempt have been made to account for the band splitting
and the field dependence of the dHvA frequencies corresponding to
the j-orbits on the Fermi surface (the necks).

{\it Band splitting and the dHvA experiments.} At high magnetic
field (above $B_c$) the effective Zeeman splitting (\ref{Beff})
(with $|\boldsymbol{S}_{\parallel}|=1$) results in the spin
dependence of the energy spectrum and of the neck width. The Zeeman
splitting makes the necks thicker for one spin component and thinner
for the other, which leads to the spin-dependence of the effective
mass, as mentioned above. This spin dependence can be observed for
both (110) and (111) magnetic field directions. In particular, for
the direction $\boldsymbol{B}\parallel (111)$ the saddle point
logarithmic divergence of the effective mass is possible for only
one spin component (sign (-) in Fig. \ref{FS2Fig}). With farther
increase of magnetic field (assuming $J>0$), this spin component
ceases to contribute to the dHvA signal with this frequency at all
since the electron trajectories start to leave the (d)-sheet of the
FS. This can be experimentally verified.\cite{Endo} At
$\boldsymbol{B}\parallel (110)$ one expects similar behavior, except
the (-) spin component at this field direction, ceases to contribute
to old dHvA frequency at lower field than at
$\boldsymbol{B}\parallel (111)$. At $B<B_c$ the splitting of the
energy spectrum remains, but now $|\boldsymbol{S}_{\parallel}| =
\sqrt{1-|\boldsymbol{S}_{\perp}|^2}=B/B_c$. Although, in Ref.
[\onlinecite{HarrCeIn3}] the spin dependence (or the band splitting)
of the effective mass was not studied, it has been observed rather
definitely in CeIn$_3$ under pressure.\cite{Endo}

More remarkable effect for $\boldsymbol{B}\parallel (110)$ is that
the two signs in Eq. (\ref{Esimple}) would correspond to two
different dHvA frequencies. If one of the necks is broken (sign (+)
in Fig. \ref{FS2Fig}), one of the dHvA frequencies corresponds to
the trajectory encircling the (d)-sheet of FS. Of the utmost
importance are the dHvA experiments measuring explicitly the
frequency(ies) from the j-trajectories on the FS as a function of
the field (for $\boldsymbol{B}\parallel (111)$, i.e. along the
direction of the neck), that would confirm the band structure of the
AFM CeIn$_3$, constructed in this paper.

{\it To summarize,} by making use of the peculiar shape of the
energy spectrum of CeIn$_3$ in terms of spherical (d) and neck-like
(j) Fermi surfaces, we have constructed the full (T-B) phase diagram
for antiferromagnetism in this compound in agreement with the
experiments.\cite{HarrCeIn3} We have analyzed the interplay between
antiferromagnetism and external magnetic field at the neck positions
and semi-quantitatively explained the observed enhancement of the
electron effective mass at the so-called ''hot
spots''.\cite{HarrCeIn3} We emphasize the importance of the ''2.5''
Lifshitz phase transition between magnetic CeIn$_3$ and nonmagnetic
LaIn$_3$. It was our intention to show that the details of the Fermi
surface topology are important for CeIn$_3$, although the magnetic
quantum criticality in other heavy fermion materials may bear the
universal character. A few straightforward experiments are suggested
to verify the above ideas.

{\it Acknowledgements.} One of the authors (LPG) thanks T.Ebihara
and N.Harrison for explanations regarding the electron spectrum in
CeIn$_3$. He benefited from the fruitful discussions with Z.Fisk
about antiferromagnetism and Kondo effect in Ce-compounds. The work
was supported by NHMFL through the NSF Cooperative agreement No.
DMR-0084173 and the State of Florida, and (PG) by DOE Grant
DE-FG03-03NA00066.

\end{document}